\documentclass[aps,prc,twocolumn,superscriptaddress,showpacs]{revtex4}
\usepackage{amsmath}

\newcommand{\mean}[1]{\left\langle #1 \right\rangle} 
\newcommand{\cumul}[1]{\left\langle\!\!\left\langle #1 
   \right\rangle\!\!\right\rangle} 
\newcommand{\N}{N_{\rm evts}}

\begin{document}

\preprint{ULB-TH/02-012, Saclay-T02/036}

\title{Analysis of directed flow from elliptic flow}

\author{N. Borghini}
\email{Nicolas.Borghini@ulb.ac.be}
\affiliation{Service de Physique Th{\'e}orique, CP 225, 
Universit{\'e} Libre de Bruxelles, B-1050 Brussels, Belgium}

\author{P. M. Dinh}
\email{dinh@spht.saclay.cea.fr}

\author{J.-Y. Ollitrault}
\email{Ollitrault@cea.fr}
\altaffiliation[also at ]{L.P.N.H.E., Universit{\'e} Pierre et Marie Curie, 
4 place Jussieu, F-75252 Paris cedex 05, France}

\affiliation{Service de Physique Th{\'e}orique, CEA-Saclay, 
F-91191 Gif-sur-Yvette cedex, France}

\date{\today}

\begin{abstract}
The directed flow of particles produced in ultrarelativistic heavy 
ion collisions at SPS and RHIC is so small that 
currently available methods of analysis are at the border of 
applicability.
Standard two-particle and flow-vector methods are biased by 
large nonflow correlations. 
On the other hand, cumulants of four-particle correlations, 
which are free from this bias, are plagued by large statistical 
errors. 
Here, we present a new method based on three-particle correlations, 
which uses the property that elliptic flow is large at these energies.
This method may also be useful at intermediate energies, near 
the balance energy where directed flow vanishes. 
\end{abstract}

\pacs{25.75.Ld 25.75.Gz}

\maketitle

\section{Introduction}

The azimuthal angles of particles emitted in a heavy  ion collisions 
are correlated with the direction of the impact parameter 
(reaction plane) of the two incoming nuclei. 
This correlation is characterized by the 
Fourier coefficients of the particle azimuthal distribution,
$v_n$~\cite{Voloshin:1996mz}:
\begin{equation}
\label{defvn}
v_n\equiv\mean{\cos(n(\phi - \Phi_R))},
\end{equation}
where $n$ is an integer, 
$\phi$ denotes the azimuthal angle of an outgoing particle,
$\Phi_{R}$ is the azimuth of the reaction plane, and angular brackets
denote an average over events and 
over all particles in a given transverse momentum and
rapidity window. 

The first Fourier coefficient $v_1$, referred to as {\em directed flow}, is 
usually positive in the forward hemisphere. 
This is because particles are deflected away from the target. 
This effect was first observed at Bevalac in 1984 \cite{Gustafsson:1984ka}, 
and reaches its maximum value at about 400~MeV per nucleon \cite{FOPI}. 
At much lower energies, below 100~MeV per nucleon, $v_1$ becomes negative due 
to the attractive nuclear mean field~\cite{balance,FOPI2}. 
At ultrarelativistic energies, it decreases: at the CERN Super Proton 
Synchrotron (SPS), it is of the order of 2\% \cite{NA49}. 
At the Brookhaven Relativistic Heavy Ion Collider (RHIC), it has not been 
measured yet. 
Accurate measurements of $v_1$ are important since this 
quantity is a sensitive probe of nuclear matter 
properties~\cite{theory}.
There is an interesting prediction that this quantity may
even become negative above SPS energies~\cite{Snellings:2000bt}.

Standard methods for analyzing directed 
flow~\cite{Danielewicz:1985hn,Wang:1991qh,Poskanzer:1998yz}
are based on the assumption that all azimuthal correlations between 
outgoing particles are due to flow. 
When flow becomes too small, however, other sources of correlation (``nonflow'' 
correlations) can no longer be neglected. 
Correlations due to global momentum conservation are well 
known~\cite{Danielewicz:1988in}, and a systematic method 
to correct for this effect has recently been 
proposed~\cite{Borghini:2002mv}. 
However, many other effects are expected to produce nonflow 
correlations of the same order of magnitude, such as quantum 
(HBT) correlations~\cite{Dinh:1999mn} and resonance 
decays~\cite{Borghini:2000cm}.
We have proposed new methods, based on higher order 
correlations~\cite{Borghini:2001sa,Borghini:2001vi}, 
which allow one to get rid of nonflow correlations.
However, statistical errors on four-particle (or higher order) 
correlations are larger, so that these methods may fail if flow is too small.

In this paper, we propose a new method which is free from these limitations. 
It allows one to obtain an estimate of $v_1$ which is insensitive to 
nonflow correlations, with statistical errors usually not much 
larger than with standard techniques. 
This method is based on the observation that elliptic flow ($v_2$, second 
Fourier harmonic of the azimuthal distribution) 
is reasonably large in the situations considered where directed flow 
is hard to measure: $v_2$ is roughly 3\% at SPS~\cite{NA49}, 
and even larger at RHIC~\cite{STAR}. 
Our method relies on the measurement of three-particle azimuthal
correlations, which involve both $v_1$ and $v_2$:
\begin{equation}
\label{basicidea}
\mean{e^{i(\phi_1+\phi_2-2\phi_3)}}\simeq (v_1)^2 v_2,
\end{equation}
where $\phi_1$, $\phi_2$, and $\phi_3$ denote azimuthal angles 
of three particles belonging to the same event, and the 
average runs over triplets of particles emitted in 
the collision, and over events. This equation yields
$(v_1)^2 v_2$, and one then deduces $v_1$ once $v_2$ has 
been obtained from a separate analysis.

The idea of mixing  two Fourier harmonics is not new. 
It underlies measurements of elliptic flow 
relative to the event plane determined from directed 
flow~\cite{elliptic}: in these analyses, one first obtains 
$v_1$, and then $v_2$ from an equation similar to Eq.\ (\ref{basicidea}).
Here we propose to do the other way round: first measure $v_2$, 
then obtain $v_1$ from the three-particle correlation. 

In Sec.~\ref{s:principle}, we present in more detail the principle 
of the method, and we explain why it is better than other methods 
when directed flow is small. 
Then, in Sec.~\ref{s:implementation}, we describe the practical 
implementation of the method. This implementation 
makes use of two techniques already 
introduced in Refs.~\cite{Borghini:2001vi,Borghini:2001zr}:
first, a cumulant expansion allows us to eliminate the effects of 
azimuthal asymmetries in the detector acceptance; 
second, the formalism of generating functions provides 
an elegant way of constructing three-particle correlations
(and beyond), with a computer time which grows only linearly 
with the number of particles involved, while taking into account 
all possible particle triplets. 
A generalization to higher order cumulants is presented in 
Sec.\ \ref{s:higherorders}. 
Our results are discussed in Sec.\ \ref{s:discussion}.

\section{Principle and orders of magnitude}
\label{s:principle}

In this section, we compare three methods of analyzing 
directed flow: the standard two-particle technique, the 
four-particle cumulant, and the three-particle mixed 
correlation, Eq.~(\ref{basicidea}), which is the focus of 
this paper. 
In Sec.~\ref{s:principleA}, we recall how $v_1$ can be 
obtained from azimuthal correlations in various ways. 
In Sec.~\ref{s:principleB} and 
Sec.~\ref{s:principleC}, we estimate for each method 
the order of magnitude 
of errors due to nonflow correlations and finite statistics. 
Numerical estimates are given in Sec.~\ref{s:principleD}, 
where we show that three-particle correlations provide 
the most (if not only) reliable way of analyzing directed 
flow at ultrarelativistic energies. 
In Sec.~\ref{s:principleE}, we finally explain how to perform 
detailed measurement of $v_1$ as a function of transverse 
momentum and rapidity with this method. 

\subsection{Directed flow from azimuthal correlations}
\label{s:principleA}

In a given collision, 
the azimuth of the reaction plane $\Phi_R$ is unknown, so that 
the Fourier coefficients $v_n$ of the azimuthal distribution, 
defined in Eq.~(\ref{defvn}), can only be obtained from azimuthal 
correlations between the outgoing particles. 
The simplest way is to use correlations between two particles
\cite{Wang:1991qh}, labelled 1 and 2, belonging to the same event:
\begin{eqnarray}
\label{2partflow}
\mean{e^{i(\phi_1-\phi_2)}}&=&
\mean{e^{i(\phi_1-\Phi_R)} e^{i(\Phi_R-\phi_2)}}\cr
&=&\mean{e^{i(\phi_1-\Phi_R)}}\mean{ e^{i(\Phi_R-\phi_2)}}\cr
&=&(v_1)^2,
\end{eqnarray}
where, as in the previous equations, averages are taken 
over pairs of particles and over events. 
In going from the first to the second line, we have assumed 
that all azimuthal correlations are due to flow, i.e., that 
azimuthal angles relative to the reaction plane 
$\phi_1-\Phi_R$ and $\phi_2-\Phi_R$ are independent. 
The ``standard'' flow analysis~\cite{Danielewicz:1985hn} proceeds 
differently: one correlates
the azimuth of a particle with a ``flow vector'' obtained 
by summing over many particles: this involves a sum of 
two-particle correlations, and this method is in fact 
essentially equivalent to the two-particle technique. 

It is straightforward to generalize Eq.~(\ref{2partflow}) to 
higher order correlations, such as the four-particle correlation:
\begin{equation}
\label{4partflow}
\mean{ e^{i(\phi_1+\phi_2-\phi_3-\phi_4)}} = (v_1)^4, 
\end{equation}
where the average now involves all possible 4-uplets of particles 
belonging to the same event. 
This result will be used below. 

In this paper, we shall be concerned with yet another type of 
correlation, namely, a three-particle correlation which mixes 
the first two Fourier harmonics:
\begin{eqnarray}
\label{3partflow}
\mean{e^{i(\phi_1+\phi_2-2\phi_3)}}&=&
\mean{e^{i(\phi_1-\Phi_R)} e^{i(\phi_2-\Phi_R)}
e^{2i(\Phi_R-\phi_3)}}
\cr
&=&(v_1)^2 v_2.
\end{eqnarray}
If $v_2$ is already known from a previous analysis, using the 
three-particle correlation Eq.\ (\ref{3partflow}) allows one to 
extract an estimate of $v_1$. 

\subsection{Nonflow correlations}
\label{s:principleB}

The above estimates were derived under the assumption that 
all azimuthal correlations are due to flow. However, there 
are also other, ``nonflow,'' contributions to azimuthal correlations
due to various effects~\cite{Poskanzer:1998yz,Dinh:1999mn,Borghini:2000cm}.
Let us recall the order of magnitude of these correlations by 
taking a simple example: if a resonance decays into $k$ particles, 
these $k$ particles will be strongly correlated by the decay 
kinematics. 
Now, the probability that $k$ arbitrary particles seen in a 
detector originate from the same resonance scales like the 
total multiplicity $M$ of the event like $1/M^{k-1}$. 
This is generally the order of magnitude 
of the genuine $k$-particle correlation due 
to nonflow effects. 
In particular,  the two-particle nonflow correlation is of the order of
$1/M$, so that Eq.~(\ref{2partflow}) reads in fact 
\begin{equation}
\label{2partnonflow}
\mean{e^{i(\phi_1-\phi_2)}}=(v_1)^2+{\cal O}\left(\frac{1}{M}\right). 
\end{equation}
When $v_1$ becomes smaller than $1/\sqrt{M}$, the resulting error 
on $v_1$ due to the unknown nonflow term  
becomes as large as $v_1$ itself. 
This is probably the case at SPS, as shown in Ref.\ \cite{Borghini:2000cm}, 
and {\it a fortiori\/} at RHIC energies. 

A similar reasoning applies to higher order correlations. 
In the case of the four-particle correlation, Eq.~(\ref{4partflow}), 
it may for instance happen that two pions labelled 
$1$ and $3$ originate from the same $\rho$ meson, while
$2$ and $4$ come from another $\rho$ meson. This gives a 
four-particle correlation of order ${\cal O}(1/M)^2$. 
However, these pairwise correlations can be subtracted from the 
measured four-particle correlation so as to isolate the genuine 
four-particle correlation. This is the principle of the 
cumulant expansion which was proposed in 
Refs.\ \cite{Borghini:2001sa,Borghini:2001vi} to get rid of nonflow 
correlations in the flow analysis. 
The cumulant of the four-particle correlation is defined as 
\begin{widetext}
\begin{equation}
\label{defcumul4}
\cumul{e^{i(\phi_1+\phi_2-\phi_3-\phi_4)}}\equiv
\mean{e^{i(\phi_1+\phi_2-\phi_3-\phi_4)}}-
\mean{e^{i(\phi_1-\phi_3)}}\mean{e^{i(\phi_2-\phi_4)}}-
\mean{e^{i(\phi_1-\phi_4)}}\mean{e^{i(\phi_2-\phi_3)}}.
\end{equation}
\end{widetext}
According to the above discussion, the contribution of nonflow 
effects to this genuine four-particle correlation is of order 
$1/M^3$, much smaller than $1/M^2$. 
The contribution of flow follows in a straightforward way from 
Eqs.~(\ref{2partflow}) and (\ref{4partflow}), so that one may finally 
write 
\begin{equation}
\label{4partnonflow}
\cumul{e^{i(\phi_1+\phi_2-\phi_3-\phi_4)}}=-(v_1)^4
+{\cal O}\left(\frac{1}{M^3}\right).
\end{equation}

Finally, let us estimate nonflow contributions to the 
mixed three-particle correlation, Eq.~(\ref{3partflow}). 
Unlike the four-particle correlation Eq.\ (\ref{4partflow}), this 
quantity does not receive any contribution from two-particle 
correlations, since quantities such as
$\mean{e^{i(\phi_1+\phi_2)}}$ or 
$\mean{e^{i(\phi_1-2\phi_3)}}$ vanish by symmetry. 
The only nonflow correlation is the genuine three-particle 
correlation, of order $1/M^2$, and Eq.~(\ref{3partflow}) 
becomes
\begin{equation}
\label{3partnonflow}
\mean{e^{i(\phi_1+\phi_2-2\phi_3)}}
=(v_1)^2 v_2+ {\cal O}\left(\frac{1}{M^2}\right).
\end{equation}

In the following, we shall denote by 
$v_1\{2\}$, $v_1\{4\}$ and $v_1\{3\}$ the estimates of $v_1$ 
obtained from Eqs.~(\ref{2partnonflow}), (\ref{4partnonflow})
and (\ref{3partnonflow}), respectively, ignoring the 
nonflow term. 
Using these equations, one finds that the differences due to nonflow 
correlations between these estimates and the exact value $v_1$ are of order:
\begin{eqnarray}
\label{errorsnonflow}
v_1\{2\}-v_1&=& {\cal O}\left(\frac{1}{Mv_1}\right)\cr
v_1\{4\}-v_1&=& {\cal O}\left(\frac{1}{(Mv_1)^3}\right)\cr
v_1\{3\}-v_1&=& {\cal O}\left(\frac{1}{(Mv_1)(Mv_2)}\right).
\end{eqnarray}
As explained in Ref.\ \cite{Borghini:2001sa}, it is possible 
to measure $v_n$ only if $M v_n \gg 1$, and we assume 
throughout this paper that this condition holds both 
for $v_1$ and $v_2$. 
Then, Eq.~(\ref{errorsnonflow}) shows that 
estimates of $v_1$ from three- or four-particle 
correlations are much less biased by nonflow correlations 
than standard estimates from two-particle correlations.

\subsection{Statistical errors}
\label{s:principleC}

In practice, the use of higher order correlations is 
limited by statistical errors due to limited statistics. 
The various correlations encountered are quantities 
of the type $(1/N)\sum_{j=1}^{N}\cos(\Delta\phi_j)$, 
where the $\Delta\phi_j$ are various combinations of the 
particle azimuths. If the $\Delta\phi_j$ are independent
and randomly distributed, the standard error on such a 
quantity is $1/\sqrt{2N}$. 

In the case of two-particle correlations, Eq.~(\ref{2partflow}), 
one can construct $M(M-1)/2$ different pairs of particles 
in an event with multiplicity $M$, so that the total number 
of combinations is $N\simeq \N M^2/2$ for $\N$ events. 
In the case of four-particle correlations, Eq.~(\ref{4partflow}), 
one can construct $M(M-1)(M-2)(M-3)/8$ independent 4-uplets, 
so that $N\simeq \N M^4/8$. 
Finally, in the case of three-particle correlations, 
there are $M(M-1)(M-2)/2$ independent triplets, so that 
$N\simeq \N M^3/2$. 

One thus obtains the following expressions for the 
relative statistical errors on the various estimates of $v_1$:
\begin{eqnarray}
\label{errorstat}
\frac{\delta v_1\{2\}}{v_1}&\simeq &\frac{1}{2\sqrt{\N}}
\,\frac{1}{(v_1\sqrt{M})^2}\cr
\frac{\delta v_1\{4\}}{v_1}&\simeq &\frac{1}{2\sqrt{\N}}
\,\frac{1}{(v_1\sqrt{M})^4}\cr
\frac{\delta v_1\{3\}}{v_1}&\simeq &\frac{1}{2\sqrt{\N}}
\,\frac{1}{(v_1\sqrt{M})^2 (v_2\sqrt{M})}.
\end{eqnarray}
These estimates are correct as long as $v_1\sqrt{M}$ and 
$v_2\sqrt{M}$ are not larger than unity, otherwise 
correlations due to flow must be taken into account in the
calculation of statistical errors~\cite{Borghini:2001vi}.
A more accurate formula for $\delta v_1\{3\}$ will be given below in 
Sec.~\ref{s:errors}. 

In this paper, we are interested in the situation where 
directed flow is very small, $v_1\sqrt{M}\ll 1$, 
which is the case when the standard flow analysis fails 
due to large nonflow correlations (see Sec.~\ref{s:principleB}). 
In that case, one sees that the statistical uncertainty on 
$v_1\{4\}$, i.e., the estimate from the fourth-order cumulant, 
is much larger than the error on the standard estimate $v_1\{2\}$. 
On the other hand, if elliptic flow is significantly large, 
$v_2\sqrt{M}$ is not much lower than unity: in such a case, the 
mixed-correlation technique provides an estimate of $v_1$
with a statistical error of the same order as the standard 
analysis, but which is much less biased by nonflow correlations.

\subsection{Orders of magnitude at SPS and RHIC}
\label{s:principleD}

Let us estimate numerically the various errors 
discussed above in a realistic situation. 
In practice, evaluating the magnitude of nonflow 
correlations requires a detailed modeling. 
Equations~(\ref{errorsnonflow}), strictly speaking,  
represent scaling laws rather than orders of magnitude. 
Detailed studies of various 
effects at SPS energies~\cite{Borghini:2000cm,Dinh:1999mn,Borghini:2002mv}
show that the contribution of nonflow correlations 
to $v_1$ obtained from the standard analysis (which is equivalent
to $v_1\{2\}$ above) is of the same order as $v_1$ itself. 
For mid-central collisions, $v_2$ is about $3\%$~\cite{NA49}
and the number of detected particles $M\sim 300$. 
According to Eq.~(\ref{errorsnonflow}), one expects 
that using three-particle correlations (i.e., $v_1\{3\}$) 
will reduce systematic errors due to nonflow correlations by 
a factor of at least $Mv_2$, that is, a factor of 10. 
Thus one expects estimates using higher order correlations 
$v_1\{3\}$ and $v_1\{4\}$ to be little biased by 
nonflow correlations, contrarily to $v_1\{2\}$. 

Statistical errors can be estimated 
quantitatively using Eqs.~(\ref{errorstat}). 
Realistic values at SPS are $\N=50$k events, $M=300$ particles, 
$v_1\simeq 2\%$ and $v_2\simeq 3\%$. 
One then obtains 
\begin{eqnarray} 
\label{errorstatSPS}    
\frac{\delta v_1\{2\}}{v_1}&\simeq &2\%\cr
\frac{\delta v_1\{4\}}{v_1}&\simeq &16\% \cr
\frac{\delta v_1\{3\}}{v_1}&\simeq &4\%.
\end{eqnarray}  
The relative statistical error on $v_1\{4\}$ is too large 
to allow detailed measurements of $v_1$ as a function of $p_T$
or $y$. On the other hand, the uncertainty on $v_1\{3\}$ 
is only twice larger than the error on $v_1\{2\}$, 
while we have seen above that the gain on the systematic error due to 
nonflow correlations is typically a factor of 10. 

At RHIC, $v_2$ is larger, typically 5\%  for mid-central 
collisions \cite{STAR}, so that measuring 
directed flow from elliptic flow ($v_1\{3\}$) is even more appropriate
than at SPS. This is reflected in both systematic errors 
due to nonflow correlations, and in statistical errors.  
While $v_2$ is larger than at SPS, $v_1$ is expected 
to be smaller if one extrapolates the decrease observed 
at SPS compared to AGS. 
The number of detected particles $M$ can be estimated using 
the values of $v_2$ and the event plane resolution given in 
Ref.\ \cite{STAR}, and is similar to that used 
above for SPS, $M\simeq 300$. 

Since $v_1$ is smaller than at SPS, 
one expects from Eq.~(\ref{errorsnonflow}) that the 
bias on $v_1\{2\}$ from nonflow correlations will be even worse. 
However, the decrease in $v_1$ may be partially compensated 
by the increase in $v_2$, so that the error on $v_1\{3\}$ 
remains of the same order. 
With $\N=50$k events, $v_1=1\%$, $v_2=5\%$, one obtains 
the following statistical errors 
\begin{eqnarray} 
\label{errorstatRHIC}   
\frac{\delta v_1\{2\}}{v_1}&\simeq &7\%\cr
\frac{\delta v_1\{4\}}{v_1}&\simeq &250\% \cr
\frac{\delta v_1\{3\}}{v_1}&\simeq &11\%,
\end{eqnarray}  
where we have used the formula derived in Sec.~\ref{s:errors}
for $\delta v_1\{3\}$ 
[here, $v_2\sqrt{M}$ is of order unity, so that 
the third of Eqs.~(\ref{errorstat}) no longer applies.]  
The statistical uncertainty is only $50\%$ larger on $v_1\{3\}$ 
than on $v_1\{2\}$, while once again the gain on the systematic 
error more than compensates for that loss. 

These numerical estimates show clearly that three-particle mixed 
correlations offer the best compromise to measure $v_1$  
at ultrarelativistic energies. 
The corresponding estimate $v_1\{3\}$ is much less affected by 
nonflow correlations than standard two-particle methods. 
In this respect, it shares the advantages of higher order 
estimates such as $v_1\{4\}$. 
In addition, $v_1\{3\}$ is much less limited by statistics 
than the latter, which would require millions of events at 
SPS and RHIC energies. 

\subsection{Integrated flow, differential flow}
\label{s:principleE}

As with other methods, the analysis proceeds in two steps. 
One first estimates the average value of $v_1$ over phase space. 
This is done by averaging $e^{i(\phi_1+\phi_2-2\phi_3)}$ over all possible 
triplets, as in Eq.~(\ref{basicidea}). 
However, $v_1$ and $v_2$ depend strongly on rapidity and transverse
momentum (for instance, $v_1$ has opposite signs in the 
backward and forward hemispheres). In practice, one therefore performs 
a {\em weighted} average, and Eq.~(\ref{basicidea}) becomes 
\begin{equation}
\label{integrated}
\mean{w_1(1)w_1(2)w_2(3)e^{i(\phi_1+\phi_2-2\phi_3)}}=
\mean{w_1v_1}^2 \mean{w_2v_2},
\end{equation}
where $w_1$ and $w_2$ are weights appropriate to directed flow 
and elliptic flow, respectively, 
which may be any function of the particle type, 
its transverse momentum $p_T$ and rapidity $y$.
In this equation, $w_n(k)$ is a shorthand for $w_n({p_T}_k,y_k)$. 
The right-hand side (rhs) of Eq.\ (\ref{integrated}) 
naturally involves {\em weighted} averages 
$\mean{w_n v_n}$, rather than $v_n$. 

The best choice for the weights is that which leads to the 
smallest statistical errors. Repeating the discussion in
Sec.~\ref{s:principleC}, one easily shows that this is done 
by maximizing $\mean{w_nv_n}/\sqrt{\mean{w_n^2}}$. 
Therefore the best weight is the flow itself 
\cite{Danielewicz:1995,Borghini:2001sa}, 
$w_n(p_T,y) = v_n(p_T,y)$, where $v_n(p_t,y)$ denotes the value of the flow in 
a small $(p_T,y)$ bin. 
In practice, one can choose as a first guess the center-of-mass 
rapidity for directed flow, $w_1=y-y_{\rm CM}$, 
and the transverse momentum for elliptic flow $w_2=p_T$, 
in regions of phase space covered by the detector acceptance. 

Using the value of the integrated (and weighted) elliptic flow 
$\mean{w_2v_2}$ obtained from a separate analysis, one finally obtains 
the integrated directed flow $\mean{w_1v_1}$ from Eq.~(\ref{integrated}). 
Naturally, one must use the same weight $w_2$ in the reference analysis 
which gives $\mean{w_2v_2}$ and in the mixed-correlation analysis. 

The second step is to analyze differential flow, i.e., to obtain values of 
$v_1$ as a function of transverse momentum $p_T$ and/or rapidity $y$. 
For that purpose, one averages $e^{i(\phi_1+\phi_2-2\phi_3)}$ over all 
$\phi_2$ and $\phi_3$, but restricts $\phi_1$ to a given particle type in a 
particular $(p_T,y)$ bin. 
\begin{equation}
\label{differential}
\mean{w_1(2)w_2(3)e^{i(\phi_1+\phi_2-2\phi_3)}}=
\mean{w_1v_1}\mean{w_2v_2} v_1(p_T,y).
\end{equation}
Note that there is no weight for the ``differential'' particle
labelled 1, for which we do not perform any phase space average. 
With the previously derived values of $\mean{w_2v_2}$ and 
$\mean{w_1v_1}$, one obtains the differential flow $v_1(p_T,y)$. 
This differential flow $v_1(p_T,y)$ will be denoted by $v'_1$ in this paper, 
while we keep the notation $v_1$ for the integrated value only. 

\section{Implementation}
\label{s:implementation}

In this section, we show how to analyze directed flow in practice 
using the three-particle correlation method. 
The method proposed here is a straightforward generalization of the one 
introduced in Ref.\ \cite{Borghini:2001vi}, which involves the formalism 
of cumulants and generating functions. 
One may believe at first sight that this formalism is a
useless complication here. But in fact, it provides an
efficient and elegant solution to the following 
problems:
\begin{itemize}
\item{Taking into account azimuthal asymmetries in the detector 
acceptance, which always exist, even if the detector has 
full azimuthal coverage.}
\item{Eliminating correlations due to detector effects.}
\item{Dealing with the combinatorics, i.e., averaging over all possible
triplets.}
\end{itemize}
We define the cumulants in Sec.~\ \ref{s:cumulants}.
The generating functions used in analyzing 
integrated and differential flow are introduced in 
Sec.\ \ref{s:genfunc}. 
Then we give interpolation formulas which can be used to extract the relevant 
cumulants from these generating functions (Sec.\ \ref{s:interpolation}). 
Finally, in Sec.\ \ref{s:errors}, we derive the standard statistical errors 
on both integrated and differential flow.

\subsection{Cumulants}
\label{s:cumulants}

Even if the detector has full azimuthal coverage, 
its acceptance is not perfectly isotropic, so that 
averages like $\mean{e^{i\phi}}$ do not strictly 
vanish. 
A general way to take into account such effects consists in 
using cumulants. For instance, the cumulant associated 
with the two-particle correlation Eq.\ (\ref{2partflow}) is defined as 
\begin{equation}
\label{cumul2}
\cumul{e^{i(\phi_1-\phi_2)}}\equiv
\mean{e^{i(\phi_1-\phi_2)}}-\mean{e^{i\phi_1}}\mean{e^{-i\phi_2}}.
\end{equation}
If the detector is perfectly isotropic, the last term vanishes 
and the cumulant reduces to the two-particle correlation. 
With a realistic detector, however, the cumulant 
isolates the physical correlation 
by subtracting the contribution of detector effects.
Similarly, the cumulant of the three-particle correlation 
is defined as 
\begin{eqnarray}
\label{cumul3}
\cumul{e^{i(\phi_1+\phi_2-2\phi_3)}} & \equiv & 
\mean{e^{i(\phi_1+\phi_2-2\phi_3)}} \cr 
& &-\mean{e^{i(\phi_1+\phi_2)}}\mean{e^{-2i\phi_3}} \cr 
& &-\mean{e^{i\phi_1}} \mean{e^{i(\phi_2-2\phi_3)}} \cr
& &-\mean{e^{i\phi_2}} \mean{e^{i(\phi_1-2\phi_3)}} \cr
& &+2\mean{e^{i\phi_1}} \mean{e^{i\phi_2}}\mean{e^{-2i\phi_3}}. \qquad
\end{eqnarray}
As in the previous case, this isolates the genuine three-particle 
correlation from effects of detector inefficiencies, and from 
spurious correlations induced by detector effects. 

If the acceptance is almost azimuthally symmetric, then all
acceptance corrections are taken into account automatically 
by the cumulant expansion: in order to obtain the flow, 
one simply need replace 
the left-hand side (lhs) of Eq.~(\ref{3partflow}) by the cumulant 
Eq.\ (\ref{cumul3}). 
If detector asymmetries are stronger, a multiplicative factor 
appears in the rhs of Eq.~(\ref{3partflow}) relating 
the correlation to the flow. 
This correction is derived in appendix \ref{s:acceptance}.

\subsection{Generating functions}
\label{s:genfunc}

Generating functions provide an elegant way of summing over 
all possible $n$-uplets in a given event. 
For a given event with $M$ particles seen in the detector, 
we define the following real-valued function of two complex variables 
$z_1=x_1+iy_1$ and $z_2=x_2+iy_2$,
\begin{widetext}
\begin{eqnarray}
\label{newG0}
G(z_1, z_2) &=& \prod_{j=1}^M 
\left[ 1 +\frac{w_{1}(j)}{M}\left( z_1^* e^{i\phi_j} + z_1e^{-i\phi_j}\right) + 
\frac{w_{2}(j)}{M}\left( z_2^* e^{2i\phi_j} + z_2e^{-2i\phi_j}\right) \right] \cr
 &=& \prod_{j=1}^M \left[ 
1 +\frac{w_{1}(j)}{M}\left( 2\,x_1 \cos(\phi_j) + 2\,y_1\sin(\phi_j)\right) + 
\frac{w_{2}(j)}{M}\left( 2\,x_2\cos(2\phi_j) + 2\,y_2\sin(2\phi_j)\right) \right], 
\end{eqnarray}
\end{widetext}
where $z_1^*\equiv x_1-iy_1$ and $z_2^*\equiv x_2-iy_2$ are the complex 
conjugates of $z_1$ and $z_2$, respectively, and $w_{1}(j)$ and 
$w_{2}(j)$ are the weights mentioned in Sec.\ \ref{s:principleE}. 
For sake of simplicity, we drop these weights from now on, unless 
otherwise stated. 

The generating function $G(z_1,z_2)$ 
generalizes the generating function $G_n(z)$ introduced in
Ref.~\cite{Borghini:2001vi}: the latter involved only one Fourier 
harmonic at a time, while we are now mixing two Fourier harmonics. 
More specifically, we recover the results of our earlier 
work in the limiting cases when either $z_1$ or $z_2$ vanishes:  
$G(z,0)=G_1(z)$, $G(0,z)=G_2(z)$.

Neither the generating function $G(z_1, z_2)$, nor the complex 
numbers $z_1$, $z_2$ have a physical meaning. 
They are a formal trick which allows us to extract azimuthal 
correlations to all orders, if necessary. 
This is done by averaging $G(z_1,z_2)$ over events 
(we denote this average by $\mean{G(z_1,z_2)}$), and then 
expanding in power series of $z_1$, $z_1^*$, $z_2$ and $z_2^*$. 
For instance, the coefficient of ${z_1^*}^2 z_2$ in the expansion is 
\begin{equation}
\label{defG}
\mean{G(z_1, z_2)}= \cdots + \frac{{z_1^*}^2 z_2}{M^3}
\mean{ \sum_{j,k,l} e^{i(\phi_j+\phi_k-2\phi_l)}} + \cdots.
\end{equation}
The sum runs over nonequivalent triplets, i.e., $j<k$. 
The values of ($j$, $k$, $l$) are all 
different, so that autocorrelations are automatically avoided. 
Since there are $M(M-1)(M-2)/2$ nonequivalent triplets, one obtains 
for large $M$ 
\begin{equation}
\mean{G(z_1, z_2)}= \cdots + \frac{{z_1^*}^2 z_2}{2}
\mean{ e^{i(\phi_1+\phi_2-2\phi_3)}} + \cdots.
\end{equation}
One recognizes here the three-particle correlation Eq.\ (\ref{basicidea}), 
averaged over triplets of particles and over events. 

In our averaging over events, we have assumed that the number of particles
$M$ is the same for all events. 
Although the method can accommodate small fluctuations of the multiplicity $M$, 
this is a possible source of error~\cite{Borghini:2001vi}. 
Our recommendation is the following: in a given centrality bin, where 
the total number of detected particles $M_{\rm tot}$ fluctuates
from one event to the other, choose a {\em fixed} number of particles 
$M\leq M_{\rm tot}$ to construct the generating function Eq.\ (\ref{newG0}). 
Although using only a fraction of the total number of particles results in a 
loss in statistics, this avoids the uncontrolled effects due to fluctuations 
in the multiplicity. 

Note that the generating function previously introduced in 
Refs.\ \cite{Borghini:2001vi,Borghini:2001zr} was a function of 
only one complex variable $z$. Here, we need two independent 
variables $z_1$ and $z_2$ because we are mixing two different 
Fourier harmonics of the azimuthal distribution. 

Generating functions are not only a convenient way of 
summing over all $n$-uplets of particles. 
They also allow one to construct easily cumulants of arbitrary order. 
The generating function of cumulants for {\em integrated} flow, 
${\cal C}(z_1,z_2)$, is defined by \cite{Borghini:2001vi,Borghini:2001zr}
\begin{equation}
\label{defC}
{\cal C}(z_1,z_2) \equiv M \left( \mean{G(z_1, z_2)}^{1/M} - 1 \right). 
\end{equation}
Expanding in power series of $z_1$, $z_1^*$, $z_2$ and $z_2^*$, one 
obtains cumulants of arbitrary order. 
In particular, the coefficient of ${z_1^*}^2 z_2$ in the expansion is 
\begin{equation}
\label{defC3}
\mean{{\cal C}(z_1, z_2)}= \cdots + \frac{{z_1^*}^2 z_2}{2}
\cumul{ e^{i(\phi_1+\phi_2-2\phi_3)}} + \cdots 
\end{equation}
An explicit calculation using Eqs.\ (\ref{newG0}) and (\ref{defC}) 
shows that the expression of 
$\cumul{e^{i(\phi_1+\phi_2-2\phi_3)}}$ thus defined coincides 
with Eq.~(\ref{cumul3}) in the limit of large $M$. 

While the generating function ${\cal C}(z_1,z_2)$ is real-valued, 
the cumulant $\cumul{e^{i(\phi_1+\phi_2-2\phi_3)}}$ defined 
by Eq.\ (\ref{defC3}) is in general a complex number. 
However, the imaginary part results from detector effects and statistical 
fluctuations, and only the real part is relevant. 
For sake of brevity, we denote this cumulant of three-particle 
correlations by $c\{3\}$ in the following: 
\begin{equation}
\label{defc1(3)}
c\{3\} \equiv {\rm Re} \left(\cumul{e^{i(\phi_1+\phi_2-2\phi_3)}} \right), 
\end{equation}
where ${\rm Re}$ denotes the real part. 
Restoring the weights, this cumulant 
gives an estimate of the weighted integrated directed flow, 
which we denote by $\mean{w_1v_1\{3\}}$ [see Eq.~(\ref{integrated})]:
\begin{equation}
\label{c3&flow}
c\{3\} = \mean{w_1v_1\{3\}}^2\,\mean{w_2v_2}, 
\end{equation}
where the integrated elliptic flow  $\mean{w_2v_2}$ 
comes from an independent analysis. 

Let us now turn to differential flow. 
We shall denote by $\psi$ the azimuth of the differential particle under study, 
and by $v'_1$ its flow, $v'_1\equiv\mean{ e^{i(\psi-\Phi_R)}}$, and call it a 
``proton'' (although it can be any type of particle). 
In opposition, we call ``pions'' the particles used to estimate integrated flow. 

The overall procedure in the analysis is quite similar to the analysis of 
integrated flow. 
We first introduce a generating function of the azimuthal correlations between 
the proton and the pions. 
It is given by the average value over {\em protons} of $e^{i\psi} G(z_1, z_2)$, 
where $G(z_1,z_2)$ is evaluated for the event where the protons belong: 
\begin{equation}
\mean{e^{i\psi} G(z_1, z_2)} = 
\mean{e^{i\psi}} + z_1 \mean{e^{i(\psi-\phi_1)}} + \ldots.
\end{equation}
Note that in the averaging procedure, an event with two protons is counted 
twice, while an event with no proton does not contribute. 

Then we define a generating function of the cumulants for {\em differential} 
flow, ${\cal D}(z_1,z_2)$, by \cite{Borghini:2001vi,Borghini:2001zr}
\begin{equation}
\label{defD}
{\cal D}(z_1,z_2) \equiv 
\frac{\mean{e^{i\psi}\,G(z_1, z_2)}}{\mean{G(z_1, z_2)}}, 
\end{equation}
where, in the denominator, $\mean{G(z_1, z_2)}$ denotes an average over 
{\em all} events. 
The cumulants are the coefficients in the expansion in power series of this 
generating function. 
As in the case of integrated flow, they are in general complex numbers, but 
only the real part is relevant from the physical point of view. 
For instance, the coefficient of $z_1^* z_2$ defines the cumulant of order 3 
which we shall use to extract differential flow: 
\begin{equation}
\label{defd1(3)}
d\{3\} \equiv {\rm Re} \left(\cumul{e^{i(\psi+\phi_1-2\phi_2)}} \right). 
\end{equation}
Note the similarity between this expression and Eq.\ (\ref{defc1(3)}). 
In fact, the cumulant $d\{3\}$ shares the same features as $c\{3\}$: it is 
free from detector effects, and reflects physical three-particle correlations, 
due either to direct three-body correlations, or to flow. 
Restoring the weights, the relation between the cumulant and flow reads
[see Eq.~(\ref{differential})]:
\begin{equation}
\label{d3&flow}
d\{3\} = \mean{w_1v_1}\mean{w_2v_2}v'_1\{3\}.
\end{equation}
We denote by $v'_1\{3\}$ this estimate of $v'_1$ obtained from three-particle 
correlations.

\subsection{Interpolating the cumulants}
\label{s:interpolation}

In this section, we show how to extract the cumulants for integrated and 
differential flows, Eqs.\ (\ref{defc1(3)}) and (\ref{defd1(3)}), numerically, 
from the computation of the generating function Eq.\ (\ref{newG0}) for various 
values of $z_1$ and $z_2$. 
We introduce the interpolation points 
$(z_{1,p}, z_{2,q})=(x_{1,p}+iy_{1,p}$, $x_{2,q}+iy_{2,q})$ with
\begin{eqnarray}
\label{interpoints}
& \displaystyle x_{1,p} =  r_0 \cos\left(\frac{p\pi}{8}\right), \quad
y_{1,p} =  r_0 \sin\left(\frac{p\pi}{8}\right), &\cr
& \displaystyle x_{2,q} =  r_0 \cos\left(\frac{q\pi}{4}\right), \quad
y_{2,q} =  r_0 \sin\left(\frac{q\pi}{4}\right), &
\end{eqnarray}
for $p=0,\ldots, 7$ and $q=0, \ldots, 3$, and
where $r_0$ is a real number, which must be neither too large, otherwise the 
error due to higher order terms in the power-series expansion of $G(z_1,z_2)$ 
is large, nor too small, to avoid numerical errors. 

To obtain the cumulants, one should for each event choose randomly $M$ 
particles among the $M_{\rm tot}$ detected, and, with the particle azimuths 
$\phi_j$ (and possibly with their transverse momenta and rapidities, if $p_T$ 
and/or $y$-dependent weights are used) compute the generating function, 
Eq.\ (\ref{newG0}), at the points, Eq.\ (\ref{interpoints}). 
Then one must average the values $G(z_{1,p},z_{2,q})$ over events, and 
calculate the generating function of cumulants, Eq.\ (\ref{defC}). 
We denote by $C_{p,q}$ the values of the generating function ${\cal C}(z_1,z_2)$ 
evaluated at the interpolation points (\ref{interpoints}):
\begin{equation}
C_{p,q} \equiv {\cal C}\left(z_{1,p}, z_{2,q}\right).
\end{equation}
From this quantities, we then build
\begin{eqnarray}
(C_p)_x &\equiv& 
\displaystyle \frac{1}{4\,r_0} \left( C_{p,0}-C_{p,2} \right), \cr
(C_p)_y &\equiv& 
\displaystyle \frac{1}{4\,r_0} \left( C_{p,3}-C_{p,1} \right), 
\end{eqnarray}
which correspond to the real and imaginary parts of the derivative of 
${\cal C}(z_1,z_2)$ with respect to $z_2$. 
Finally, the third order cumulant we are interested in, $c\{3\}$, is given by
\begin{eqnarray}
\label{interpolc1(3)}
c\{3\} &=& \displaystyle 
\frac{1}{4\,r_0^2} \left[(C_0)_x-(C_1)_y-(C_2)_x+(C_3)_y\right. \cr
 & & \qquad \displaystyle\left. +\ (C_4)_x-(C_5)_y-(C_6)_x+(C_7)_y \right].\quad
\end{eqnarray}

Consider now differential flow. 
We denote by $D_{p,q}$ the values of the generating function ${\cal D}(z_1,z_2)$ 
evaluated at the interpolation points (\ref{interpoints}):
\begin{equation}
\label{Dpq}
D_{p,q} \equiv (D_{p,q})_x+i(D_{p,q})_y \equiv 
{\cal D}\left(z_{1,p}, z_{2,q}\right), 
\end{equation}
where in fact one only need use even values of $p$: the interpolation of the 
``differential'' cumulant, which is only a second order derivative, requires 
less points than the ``integrated'' cumulant, which is a derivative of third 
order. 
With the quantities Eq.\ (\ref{Dpq}), we then build
\begin{eqnarray*}
(D_p)_x &\equiv& \displaystyle \frac{1}{4\,r_0} 
\left[ (D_{p,0})_x - (D_{p,2})_x + (D_{p,1})_y - (D_{p,3})_y \right], \cr
(D_p)_y &\equiv& \displaystyle \frac{1}{4\,r_0} 
\left[ (D_{p,0})_y - (D_{p,2})_y + (D_{p,3})_x - (D_{p,1})_x \right],
\end{eqnarray*}
and the cumulant $d\{3\}$ is finally given by
\begin{equation}
\label{interpold1(3)}
d\{3\} = \frac{1}{4\,r_0} \left[(D_0)_x-(D_2)_y-(D_4)_x+(D_6)_y\right]. 
\end{equation}

Naturally, one may prefer using a different interpolation scheme. 
In any case, one should check that the final values, $c\{3\}$ and $d\{3\}$, 
do not depend on the parameters introduced in the interpolation: here, one 
should try different values of $r_0$, and check the stability of the result.

\subsection{Statistical errors}
\label{s:errors}

The standard deviation of the cumulant $c\{3\}$, Eq.(\ref{defc1(3)}), 
is given by 
\begin{equation}
\label{deltac3}
(\delta c\{3\})^2=\frac{\mean{w_1^2}^2\mean{w_2^2}}{2M^3\N}
(2+4\chi_1^2+2\chi_2^2+4\chi_1^2\chi_2^2+\chi_1^4).
\end{equation}
In this equation, $\chi_1$ and $\chi_2$ are the resolution parameters 
appropriate for directed flow and elliptic flow, respectively:
\begin{equation}
\label{defchi}
\chi_n\equiv \frac{\mean{w_nv_n}}{\sqrt{\mean{w_n^2}}}\sqrt{M}.
\end{equation}

From Eq.~(\ref{deltac3}), and the relation between $c\{3\}$ and 
the flow, Eq.~(\ref{c3&flow}), one can calculate the statistical 
error on $\mean{w_1v_1}$. 
If one neglects the statistical error on $v_2$, 
one recovers the simple estimate (\ref{errorstat}) 
with unit weights, in the limit where $\chi_1\ll 1$ and $\chi_2\ll 1$. 
A more careful estimate must take into account the error on 
$\mean{w_2 v_2}$ in deriving the error on $\mean{w_1v_1}$.
Statistical errors on the cumulant are Gaussian. 
Since the relation between the cumulant and the integrated 
directed flow, Eq.~(\ref{c3&flow}), is quadratic rather than
linear, the resulting error bars on $\mean{w_1v_1}$ are 
asymmetric when the error is large, as discussed in detail 
in appendix D of Ref.\ \cite{Borghini:2001vi}.

Similarly, the standard deviation of the differential cumulant 
$d\{3\}$, Eq.~(\ref{defd1(3)}), is given by 
\begin{equation}
\label{deltad3}
(\delta d\{3\})^2=
\frac{\mean{w_1^2}\mean{w_2^2}}{2M^2N'}
(1+\chi_1^2)(1+\chi_2^2),
\end{equation}
where $N'$ denotes the number of protons used in the differential flow 
analysis. 
Using the relation (\ref{d3&flow}) between this cumulant and 
the differential flow, $v'_1\{3\}$, we easily obtain the statistical 
error on $v'_1\{3\}$. 
Since the analysis is done in a narrow phase space bin, one may 
reasonably assume here that the error on $d\{3\}$ 
is dominated by the error on $v'_1\{3\}$, which yields 
\begin{equation}
\delta v'_1\{3\}=\frac{1}{\sqrt{2N'}}
\frac{\sqrt{1+\chi_1^2}}{\chi_1}\frac{\sqrt{1+\chi_2^2}}{\chi_2}.
\end{equation}
This result can be understood simply in two limiting cases. 
When both $\chi_1$ and $\chi_2$ are large compared to unity, 
the reaction plane $\Phi_R$ can be 
reconstructed accurately. The error on $\mean{\cos(\psi-\Phi_R)}$ 
estimated with $N'$ values of $\psi$ is then $1/\sqrt{2N'}$, 
to which the error reduces for $\chi_1\gg 1$, $\chi_2\gg 1$.
In the opposite case $\chi_1\ll 1$, $\chi_2\ll 1$, 
comparing Eqs.~(\ref{deltac3}) and (\ref{deltad3}), one finds 
for unit weights $\delta v'_1\{3\}/\delta v_1\{3\}=\sqrt{M\N/N'}$: 
errors scale like the inverse square root of the number of 
particles involved, and the determination of the integrated flow 
$v_1$ involves a total number of $M\N$ particles while 
the differential flow $v'_1$ involves $N'$ particles. 

\section{Higher orders}
\label{s:higherorders}

In the previous section, we have studied the three-particle 
correlation, which is the lowest order nontrivial result obtained 
with the generalized generating function. Higher orders can also 
be derived in a straightforward way. 
Expanding the generating function of cumulants ${\cal C}(z_1,z_2)$, 
defined in Eq.~(\ref{defC}), 
in power series of $z_1$, $z_1^*$, $z_2$, and $z_2^*$,
yields cumulants of arbitrary order:
\begin{widetext}
\begin{equation}
\label{expC}
{\cal C}(z_1,z_2) \equiv 
\sum_{j,k,l,m} \frac{{z_1^*}^j z_1^k {z_2^*}^l z_2^m}{j!\,k!\,l!\,m!} 
\cumul{e^{i(\phi_1+\cdots+\phi_j-\phi_{j+1}-\cdots-\phi_{j+k} + 
2(\phi_{j+k+1}+\cdots+\phi_{j+k+l}-\phi_{j+k+l+1}-\cdots-\phi_{j+k+l+m}))}}.
\end{equation}
The only relevant cumulants are those which are invariant under 
a simultaneous shift of all azimuthal angles $\phi_j\to\phi_j+\alpha$. 
Other cumulants vanish except for statistical fluctuations and detector effects.
According to Eq.~(\ref{defG}), this shift is equivalent to the 
change of variables $z_1\to z_1 e^{-i\alpha}$, 
$z_2\to z_2 e^{-2i\alpha}$. 
The only terms in Eq.~(\ref{expC}) which are invariant under this 
transformation are those with $j+2l=k+2m$. 
Any of these cumulants can be used to extract the flow. 
Until now, we have explored only a few possibilities:
the case considered in Sec.~\ref{s:implementation}
is $(j,k,l,m)=(2,0,0,1)$; 
the cumulants used in Ref.\ \cite{Borghini:2001vi} to extract $v_1$ and 
$v_2$ are those with $j=k$, $l=m=0$ and those with 
$j=k=0$, $l=m$ (denoted by $c_1\{2j\}$ and $c_2\{2l\}$, respectively).

We now derive the relations between cumulants of arbitrary orders 
and the flow coefficients $v_1$ and $v_2$. 
For this purpose, 
we compute the average value of $G(z_1,z_2)$ in the presence of flow. 
We first average for a given orientation of the reaction plane $\Phi_R$. 
For an arbitrary particle, by definition of $v_n$, we may write
$\mean{e^{in\phi}|\Phi_R}=v_ne^{in\Phi_R}$. 
Replacing in Eq.~(\ref{defG}), dropping the weights for simplicity, 
and neglecting all nonflow correlations, we obtain 
\begin{eqnarray}
\label{meanG0}
\mean{G(z_1,z_2)|\Phi_R}  
&= &\left( 1+ 
\frac{z_1^*v_1\, e^{i\Phi_R}+z_1\, v_1\, e^{-i\Phi_R} + 
z_2^*v_2\, e^{2i\Phi_R}+z_2\, v_2\, e^{-2i\Phi_R}}{M} \right)^M \cr
&\simeq &
\exp\left(z_1^*v_1\, e^{i\Phi_R}+z_1\, v_1\, e^{-i\Phi_R}\right) 
\exp\left(z_2^*v_2\, e^{2i\Phi_R}+z_2\, v_2\, e^{-2i\Phi_R}\right). 
\end{eqnarray}
\end{widetext}
The next step is to average over $\Phi_R$. 
We make use of the following formula
\begin{equation}
\exp\left( z^* e^{i\phi}+z e^{-i\phi}\right)=\sum_{q=-\infty}^{+\infty}
e^{-iq\phi} \left(\frac{z}{|z|}\right)^{\!q} I_q(2|z|), 
\end{equation}
where $I_q$ is the modified Bessel function of order $q$. 
Applying this identity to each term of Eq.~(\ref{meanG0}) and 
integrating over $\Phi_R$, one obtains 
\begin{eqnarray}
\label{avphiR}
\!\!\mean{G(z_1,z_2)} &\!\!=\!\!&
\int_0^{2\pi}\mean{G(z_1,z_2)|\Phi_R}  \frac{d\Phi_R}{2\pi}\cr
&\!\!=\!\!&\!\!\sum_{q=-\infty}^{+\infty}\!\!
\left(\frac{{z_1^*}^2 z_2}{|z_1|^2 |z_2|}\right)^{\!\!q} \! \!
I_{2q}(2|z_1| v_1) I_q(2|z_2|v_2). \quad
\end{eqnarray}
In the limiting cases $z_1=0$ or $z_2=0$, only the term $q=0$ contributes 
to the sum in the rhs, and we recover $\mean{G(z_1,0)}=I_0(2|z_1|v_1)$, 
$\mean{G(0,z_2)}=I_0(2|z_2|v_2)$, already derived in
Ref.~\cite{Borghini:2001vi}. 

Expanding the generating function of cumulants, 
${\cal C}(z_1,z_2)\simeq \ln \mean{G(z_1,z_2)}$, in powers
of $z_1$, $z_1^*$, $z_2$, $z_2^*$, and identifying with 
Eq.~(\ref{expC}), one obtains the relations between the various 
cumulants and flow. 
As expected, the only nonvanishing terms 
are those which satisfy the condition $j+2l=k+2m$ derived above, 
and the corresponding cumulants are proportional to 
$(v_1)^{j+k}(v_2)^{l+m}$, with an integer multiplicative constant 
depending on the values of $j,k,l,m$. 
To order ${z_1^*}^2 z_2^* z_2^2 $, for 
instance, one obtains 
\begin{equation}
\cumul{e^{i(\phi_1+\phi_2-2\phi_3+2\phi_4-2\phi_5)}}=-(v_1)^2 (v_2)^3.
\end{equation}
If $v_2$ is measured independently, this equation yields an estimate 
of $v_1$ from 5-particle correlations, which we denote by $v_1\{5\}$.
The statistical error on $v_1\{5\}$ is not much 
larger than the error on the earlier three-particle estimate $v_1\{3\}$
if $v_2$ is large enough (more precisely, if the resolution parameter
$\chi_2\sim v_2\sqrt{M}$ is larger than unity). 
This may be useful at RHIC, not at SPS where $v_2$ is too small. 
Cumulants with $j+k>2$ in Eq.~(\ref{expC}) involve higher powers 
of $v_1$ and are useless in the situation we are interested in, 
namely small values of $v_1$. 

\section{Discussion}
\label{s:discussion}

In the previous sections, we have presented a new method for analyzing directed 
flow ($v_1$), through the help of an independent measurement of elliptic flow 
($v_2$). 
It allows one to measure both integrated and differential flow, the value of 
integrated flow being used in the differential analysis. 

Our method relies on a study of {\em three}-particle correlations. 
Unlike standard methods, based on two-particle correlations,
it is not biased by two-particle nonflow correlations, 
which are an important bias at ultrarelativistic energies. 
This can be checked experimentally by studying $v_1$ near midrapidity:
while standard two-particle estimates $v_1\{2\}$ generally do not cross zero, 
especially if they are not corrected for 
momentum conservation \cite{Borghini:2002mv}, our estimate 
$v_1\{3\}$ should naturally vanish at midrapidity, as expected. 

In experiments where analyses of directed flow are already available, 
it would be interesting to compare these 
standard estimates from two-particle correlations, $v_1\{2\}$, 
with our new estimate from three-particle correlations $v_1\{3\}$.
If they are in agreement (within statistical error bars), 
it is a good hint that they indeed coincide with the true directed flow. 
If they differ, 
it is instructive to study the centrality dependence of the product 
$M(v_1\{3\}^2-v_1\{2\}^2)$, where $M$ is the event multiplicity. 
If the difference between $v_1\{2\}$ and $v_1\{3\}$ is due to two-particle 
nonflow correlations, this product should be approximately constant 
\footnote{Similarly, the product $M(v_n\{4\}^2-v_n\{2\}^2)$ may be used to 
probe the possible difference between a four-particle flow estimate $v_n\{4\}$ 
derived with the method proposed in Ref.\ \cite{Borghini:2001vi} and an 
estimate $v_n\{2\}$ obtained with some two-particle method.} (remember that 
two-particle nonflow correlations scale as $1/M$.) 
If the product differs significantly from a constant, then another explanation 
must be looked for: the difference between $v_1\{2\}$ and $v_1\{3\}$ may be due 
to fluctuations, either fluctuations of the impact parameter within a given 
centrality class of events \cite{Poskanzer}, 
or, more interestingly, physical fluctuations of the flow event-by-event. 

The price to pay for eliminating nonflow effects is an increase
in statistical errors, compared to standard two-particle methods. 
However, this increase is moderate, a factor of 2 or less at 
SPS and RHIC. 
This new method is thus much less statistics-demanding than 
those based on correlations between four (or more) particles. 
All in all, three-particle correlations seem to be the most 
appropriate way to measure $v_1$ when it is small, and especially if $v_2$ is strong: near the balance 
energy, at SPS, RHIC,  and the forthcoming experiments at LHC.

\begin{acknowledgments}
We thank Marek Ga{\'z}dzicki, Art Poskanzer, Herbert Str{\"o}bele, 
Sergei Voloshin and Alexander Wetzler for discussions. 
N.\ B.\ acknowledges support from the ``Actions de Recherche Concert{\'e}es'' 
of ``Communaut{\'e} Fran{\c c}aise de Belgique'' and IISN-Belgium.
\end{acknowledgments}

\appendix
\section{Acceptance corrections}
\label{s:acceptance}

In the case of a realistic detector, with a nonisotropic acceptance, the 
relations between the cumulants and flow, Eqs.\ (\ref{c3&flow}) and 
(\ref{d3&flow}), no longer hold. 
In this appendix, we derive the relevant modifications taking into account 
the detector acceptance. 

In what follows, we assume that the various classes of events analyzed, for 
instance, the different centrality bins, are determined with an 
{\em independent} detector (e.g., a ZDC), which has a full azimuthal coverage 
(at least approximately). 
This is meant to make sure that the centrality assigned to a given event is not 
strongly correlated to the orientation of its reaction plane, which would bias 
the sample of events used in the flow analysis. 

To describe a detector, we introduce its acceptance/ efficiency function 
$A(j,\phi,p_T,y)$, which is the probability that a particle of type $j$ with 
azimuth $\phi$, transverse momentum $p_T$, and rapidity $y$ be detected 
\cite{Borghini:2001vi,Borghini:2001zr}. 
Obviously, $A(j,\phi,p_T,y)$ will vary from a detector to another, and a 
``perfect'' detector corresponds to $A(j,\phi,p_T,y)=1$ for every particle type 
in the whole phase space. 
In practice, $A(j,\phi,p_T,y)$ is proportional to the number of hits in a 
$(\phi, p_T, y)$ bin: to obtain its shape for a given detector and for each 
$(p_T,y)$ bin, one only has to count the number of hits in each $\phi$ bin 
while reading the data to analyze them, and in the end divide by the maximal 
number encountered. 
The Fourier coefficients of the acceptance function are defined by
\begin{subequations}
\label{def_ak}
\begin{equation}
\label{ak_pTy}
A_n(j,p_T,y) \equiv \displaystyle \int_0^{2\pi}\! A(j,\phi,p_T,y) \, 
e^{-in\phi} \, \frac{d\phi}{2\pi}. 
\end{equation}
These differential coefficients can be integrated, with appropriate weighting 
and a sum over the various types of particles used for the flow analysis, so as 
to describe the ``integrated'' acceptance of the detector: 
\begin{equation}
\label{ak}
a_n[w] = \frac{\displaystyle \sum_j \int\! 
w(j,p_T,y)\,A_n(j,p_T,y)\, dp_T\,dy}{\displaystyle \sum_j
\int \! w(j,p_T,y)\,A_0(j,p_T,y)\, dp_T\,dy}. 
\end{equation}
\end{subequations}
Note our introducing the weights $w(j,p_T,y)$, which are of course the same as 
in Eq.\ (\ref{newG0}), either $w_1(j)$ or $w_2(j)$. 
In the following, we assume that both weights are equal, so that there is only 
one set of $a_n$ coefficients. 
If two different weights are used---a weight $w_1$ which maximizes $v_1$ and 
the weight $w_2$ which was used to derive the reference $v_2$---one should keep 
track of the two different sets of coefficients $a_n[w_1]$, $a_n[w_2]$ in the 
calculation which we now sketch. 

To compute the contribution of flow to the cumulant $c\{3\}$, we follow the 
same procedure as in Ref.\ \cite{Borghini:2001vi}. 
We first average the generating function Eq.\ (\ref{newG0}) over events with 
the same azimuth of the reaction plane $\Phi_R$; then we average over $\Phi_R$. 
For simplicity, we neglect nonflow correlations in the derivation. 
Denoting by $\mean{x|\Phi_R}$ the average of a quantity $x$ for fixed $\Phi_R$, 
the average value of $e^{in\phi}$ for a fixed $\Phi_R$ is given by 
\cite{Borghini:2001vi}
\begin{equation}
\label{vn|PhiR}
\mean{e^{in\phi}|\Phi_R} = 
a_n^* + \sum_{p\neq 0} (a_{p-n}-a_p a_n^*) v_p e^{ip\Phi_R}, 
\end{equation}
where $a_n^*=a_{-n}$ is the complex conjugate of $a_n$. 
Thus, a nonisotropic acceptance mixes the various flow harmonics. 
In the case of a perfect acceptance, Eqs.\ (\ref{def_ak}) show that all 
coefficients $a_n$ vanish, except $a_0=1$, and the identity (\ref{vn|PhiR}) 
reads $\mean{e^{in\phi}|\Phi_R} = v_n e^{in\Phi_R}$, which follows in a 
straightforward way from Eq.\ (\ref{defvn}). 

Inserting the average value, Eq.\ (\ref{vn|PhiR}) for $n=1$ and $n=2$ in 
Eq.\ (\ref{newG0}), one obtains the generating function averaged over events 
with the same orientation of the reaction plane, $\mean{G(z_1,z_2)|\Phi_R}$. 
The latter must then be averaged over $\Phi_R$, then one computes the cumulants 
using Eqs.\ (\ref{defC}) and (\ref{expC}), keeping only the real part. 
In particular, the third cumulant reads 
\begin{eqnarray}
\label{c3&flowbis}
c\{3\}\!&\!=\!&\!{\rm Re}\left( (1-|a_2|^2)\left(1-|a_1|^2\right)^2 \right.\cr
& & \qquad + 2 \left[  (a_1-a^*_1a_2)^2 \left(a^*_2-{a^*_1}^2\right) \right.\cr
& & \qquad\qquad  + \left(1-|a_1|^2\right) |a_3-a_1a_2|^2 \cr
& & \qquad\qquad  + \left. \left. (a_4-a_2^2)\left(a^*_2-{a^*_1}^2\right)^2 
\right] \right) v_1^2 v_2, \qquad 
\end{eqnarray}
instead of Eq.\ (\ref{c3&flow}), where we have assumed that only $v_1$ and 
$v_2$ are nonvanishing. 
When the detector is perfect, one recovers Eq.\ (\ref{c3&flow}). 
Even if the detector is not perfect, but nevertheless does not have too bad an
acceptance, the factor in front of $v_1^2\,v_2$ in Eq.\ (\ref{c3&flowbis}) 
will remain close to 1, since the correction terms are at least quadratic in 
the $a_n$ coefficients. 
Thus, Eq.\ (\ref{c3&flow}) remains a good approximation, except for detectors 
with a very bad azimuthal coverage. 

The contribution of flow to the ``differential'' cumulant $d\{3\}$, 
Eq.\ (\ref{defd1(3)}), can be calculated along the same lines. 
In that calculation, one may assume that integrated and differential flows 
are measured using two different detectors: 
e.g., a large acceptance detector for integrated flow, and a smaller one, but 
with better particle identification or $p_T$ determination, for differential 
flow. 
We thus denote by $A'(j,\psi,p_T,y)$ the corresponding acceptance function and 
by $A'_k(j,p_T,y)$ its Fourier coefficients defined as in Eq.\ (\ref{ak_pTy}). 
The differential acceptance coefficients $a'_k$ are then defined as in 
Eq.\ (\ref{ak}), without the weights and the summation over $j$ (since one 
usually measures the differential flow of identified particles) and with the 
integration over $p_T$ and $y$ restricted to the phase-space region under 
interest (typically, one integrates over $p_T$ {\em or} $y$, so as to obtain 
$v'_1$ as a function of $y$ or $p_T$, respectively).  

The average value over protons in the numerator of Eq.\ (\ref{defD}) is then
computed in two steps, first averaging over protons detected in events with the 
same orientation of the reaction plane, then averaging over $\Phi_R$. 
The denominator of Eq.\ (\ref{defD}), $\mean{G(z_1,z_2)|\Phi_R}$, is the same 
as in the calculation of $c\{3\}$. 
Finally, $d\{3\}$ is given by the coefficient of $z_1^* z_2$ in the expansion 
in power series of ${\cal D}(z_1,z_2)$: 
\begin{eqnarray}
\label{d1(3)acc}
d\{3\} \!&\!=\!&\!
{\rm Re}\left[ (1-|a_1|^2)(1-|a_2|^2) + (a_1-a_1^* a_2)^2 {a'_2}^* \right. \cr
 & & \left. +|a_3-a_1 a_2|^2 + 
(a_2^* - {a_1^*}^2)(a_4-a_2^2) {a'_2}^* \right] v'_1v_1v_2 \cr 
& &\! +{\rm Re}\ \left[ (1-|a_1|^2)(a_3-a_1 a_2) {a'_3}^* \right. \cr
& & \quad \left. +\ (a_2^*-{a_1^*}^2)(a_1-a_1^* a_2) a'_1 \right] v'_2 v_1^2.
\end{eqnarray}
A nonisotropic acceptance will cause interference between the various 
differential flow harmonics: the measurement of directed differential flow 
$v'_1$ is perturbed by the elliptic differential flow $v'_2$. 
It is worth noting that as soon as the acceptance of the detector used for 
integrated flow is perfect, Eq.\ (\ref{d1(3)acc}) reduces to 
Eq.\ (\ref{d3&flow}).

\end{document}